# Imaging and nulling properties of sparse-aperture Fizeau interferometers

François Hénault
Institut de Planétologie et d'Astrophysique de Grenoble
Université Joseph Fourier, Centre National de la Recherche Scientifique
B.P. 53,  38041 Grenoble – France

**ABSTRACT**

In this communication are presented rigorous and approximate analytical expressions of the Point Spread Function (PSF) and Field of View (FoV) achievable by multi-aperture Fizeau interferometers, either of the imaging or nulling types. The described formalism can be helpful for dimensioning future space missions in search of habitable extra-solar planets. Herein the characteristics of PSF and FoV are derived from simple analytical expressions that are further computed numerically in order to evidence the critical role of pupil re-imaging along the interferometer arms. The formalism is also well suited to simulating pseudo-images generated by a nulling Fizeau interferometer, and numerical computations demonstrate that it is only efficient for very short baselines. Finally, two different designs improving the nulling capacities of such exoplanet observing instruments are briefly presented and discussed.

## 1  INTRODUCTION

Since the historical writings of Fizeau [1], Stéphan [2] and Michelson [3], multi-aperture optical systems and their imaging properties have been the subject of extensive literature, leading among other to the currently admitted distinction between "Fizeau" and "Michelson" types of interferometer. It is often considered today that the major difference between both concepts relies on the fact that the former obeys to a "golden rule" stating that the output pupil of the interferometer must be a scaled replica of its entrance pupil [4-5], while the latter does not. Some new multi-aperture concepts, however, have emerged during the three last decades, such as space borne, infrared nulling interferometers or telescopes dedicated to the search of extra-solar planets [6-7]. In two recent publications was described a simple Fourier optics formalism allowing to derive the basic Object-Image relationships of such systems [8] and approximate expressions of their Point Spread Functions (PSF) and achievable Field of View (FoV) [9]. The purpose of the present communication is to complete and simplify this formalism again (section 2) and to provide a deeper analysis of sparse-aperture Fizeau interferometers. Examples of applications are described in section 3, where is illustrated the critical role of pupil re-imaging along the interferometer arms. Finally, two alternative designs improving the nulling and pseudo-imaging capacities of such instrument envisioned for future space missions searching for habitable extra-solar planets are presented and discussed in section 4.

## 2  GENERAL RELATIONSHIPS

Here the formalism described in Refs. [8-9] is briefly summarized in section 2.1, before deriving the theoretical relationships applicable to the PSF (§ 2.2), maximal achievable field of view (§ 2.3). Furthermore, an approximation regarding the maximal achievable FoV in Ref. [9] has been clarified.

### 2.1  Object-Image relationship

Let us consider an optical system designed either for high-angular resolution imaging or nulling interferometry purpose, being composed of N collecting apertures and N recombining apertures. Figure 1 depicts the three main coordinate systems being used here after: they are an on-sky angular coordinates system (U,V), an entrance pupil reference frame (O,X,Y,Z) where OZ is the main optical axis, and an exit pupil reference frame (O',X',Y',Z). Let us further assume that:

1) For all indices n comprised between 1 and N, the nth collecting aperture of center $P_n$ is optically conjugated with its associated combining aperture of center $P'_n$ without any pupil aberration.

2) All collecting apertures have an identical diameter D and consequently all recombining apertures share the same diameter D'. Practically, it means that all collecting telescopes and optical trains conveying the beams from the entrance to the exit apertures are identical, which is most often the case in current interferometer facilities, either of Fizeau or Michelson types. The generic optical layout of the interferometer is depicted in Figure 2.

Let us finally define the following parameters (bold characters denoting vectors):

| | |
|---|---|
| **s** | A unit vector of direction cosines ≈ (u,v,1) directed at any point in the sky (corresponding to any point M" in the image plane), where angular coordinates u and v are considered as first-order quantities |
| **s<sub>O</sub>** | A unit vector of direction cosines ≈ ($u_O,v_O$,1) pointed at a given sky object (or an elementary angular area of it) |
| O(**s<sub>O</sub>**) | The angular brightness distribution of an extended sky object |
| $\Omega_O$, $d\Omega_O$ | The total observed FoV in terms of solid angle, and its differentiating element |
| $PSF_T$(**s**) | The PSF of an individual collecting telescope, being projected back onto the sky. For an unobstructed pupil of diameter D, this would be the classical Airy distribution equal to $|2J_1(\rho)/\rho|^2$, where $\rho = k\,D\,\|\mathbf{s}\|/2$ and $J_1$ is the type-J Bessel function at the first order |
| K | The wavenumber $2\pi/\lambda$ of the electro-magnetic field assumed to be monochromatic, and $\lambda$ is its wavelength |
| $a_n$ | The amplitude transmission factor of the $n^{th}$ interferometer arm ($1 \leq n \leq N$) |
| $\varphi_n$ | A phase-shift introduced along the $n^{th}$ interferometer arm for Optical Path Differences (OPD) compensation or nulling purposes ($1 \leq n \leq N$) |
| **OP<sub>n</sub>** | A vector defining the center $P_n$ of the $n^{th}$ sub-pupil in the entrance pupil plane P ($1 \leq n \leq N$) |
| B | The maximal baseline between any couple (n,n') of telescopes ($1 \leq n$ and $n' \leq N$) |
| **O'P'<sub>n</sub>** | Correspondingly, a vector defining the center $P'_n$ of the $n^{th}$ sub-pupil in the exit pupil plane P' ($1 \leq n \leq N$) |
| B' | The maximal baseline between any couple (n,n') of exit sub-apertures ($1 \leq n$ and $n' \leq N$) |
| m | The optical compression factor of the system, equal to $m = D'/D = F_C/F$ where F and $F_C$ respectively are the focal length of the collecting telescopes and of the relay optics (see Figure 2). |

Hence according to Refs. [1] and [9] the expression of the image I(**s**) formed by the multi-aperture optical system and projected back onto the sky writes in a first-order approximation:

$$I(\mathbf{s}) = \iint_{\mathbf{s_O} \in \Omega_O} O(\mathbf{s_O})\, PSF_T(\mathbf{s} - \mathbf{s_O}) \left| \sum_{n=1}^{N} a_n \exp[i\varphi_n] \exp[ik\xi(\mathbf{s_O},\mathbf{s})] \right|^2 d\Omega_O , \quad (1a)$$

with function $\xi(\mathbf{s_O},\mathbf{s})$ being an extra OPD term:
$$\xi(\mathbf{s_O},\mathbf{s}) = \mathbf{s_O}\,\mathbf{OP_n} - \mathbf{s}\,\mathbf{O'P'_n}/m . \quad (1b)$$

This very general Object-Image relationship can only be reduced to convolution products if certain conditions are fulfilled, which were extensively discussed in Ref. [9]. Herein the following sections only deal with some consequences on the PSF and effective FoV accessible by the whole system.

### 2.2 Point Spread Function

We can define a "generalized PSF" of the optical system by simply replacing the object brightness function O(**s<sub>O</sub>**) with the impulse Dirac distribution $\delta(\mathbf{s}-\mathbf{s_O})$ in Eq. 1a:

$$PSF_G(\mathbf{s},\mathbf{s_O}) = PSF_T(\mathbf{s} - \mathbf{s_O}) \left| \sum_{n=1}^{N} a_n \exp[i\varphi_n] \exp[ik\xi(\mathbf{s_O},\mathbf{s})] \right|^2 . \quad (2)$$

$PSF_G(\mathbf{s},\mathbf{s_O})$ presents the particularity of constantly varying with the angular location $\mathbf{s_O}$ of the sky object in the instrument FoV, hence differing significantly from the familiar, invariant PSF of Fourier optics. Consequently, the notions of

Optical Transfer Function (OTF) and Modulation Transfer Function (MTF) cannot by applied in classical sense and are not discussed further.

## 2.3 Maximal achievable Field of View

We may intuitively define a "maximal achievable Field of View" of the multi-aperture optical system as the image that would be formed under the following hypotheses:

- o No physical diaphragm of any kind (stops, mirrors edge or central obscuration) is taken into account.
- o The sky object is uniformly bright over a $2\pi$-steradian solid angle, hence $O(\mathbf{s_O}) = 1$.
- o The optical system is free from aberrations, thus $PSF_T(\mathbf{s})$ is assumed to be the classical Airy function.

Under such assumptions the maximal achievable FoV deduced from Eq. 1a is:

$$\text{FoV}(\mathbf{s}) = \iint\limits_{\mathbf{s_O} \in \Omega_O} PSF_T(\mathbf{s} \cdot \mathbf{s_O}) \left| \sum_{n=1}^{N} a_n \exp[i\varphi_n] \exp[ik\xi(\mathbf{s_O}, \mathbf{s})] \right|^2 d\Omega_O . \quad (3)$$

Eq. 3 is still a complicated mathematical expression that cannot be simplified analytically and is requiring extensive computing times when calculated numerically. An additional heuristic simplification presented in Ref. [9] consisted in assuming that $PSF_T(\mathbf{s})$ can be approximated to the Dirac distribution $\delta(\mathbf{s})$. In that case the integral of Eq. 3 can be reduced to:

$$\text{FoV}(\mathbf{s}) = \left| \sum_{n=1}^{N} a_n \exp[i\varphi_n] \exp[ik\mathbf{s}(\mathbf{OP_n} - \mathbf{O'P'_n}/m)] \right|^2 . \quad (4)$$

This is a simplified and condensed expression of the theoretical achievable FoV. However Eq. 4 remains an approximate relationship that should only be used with the greatest care, because it is only valid for when individual telescopes are of large aperture size, or they are close one to the other. In other cases Eq. 3 has to be computed accurately.

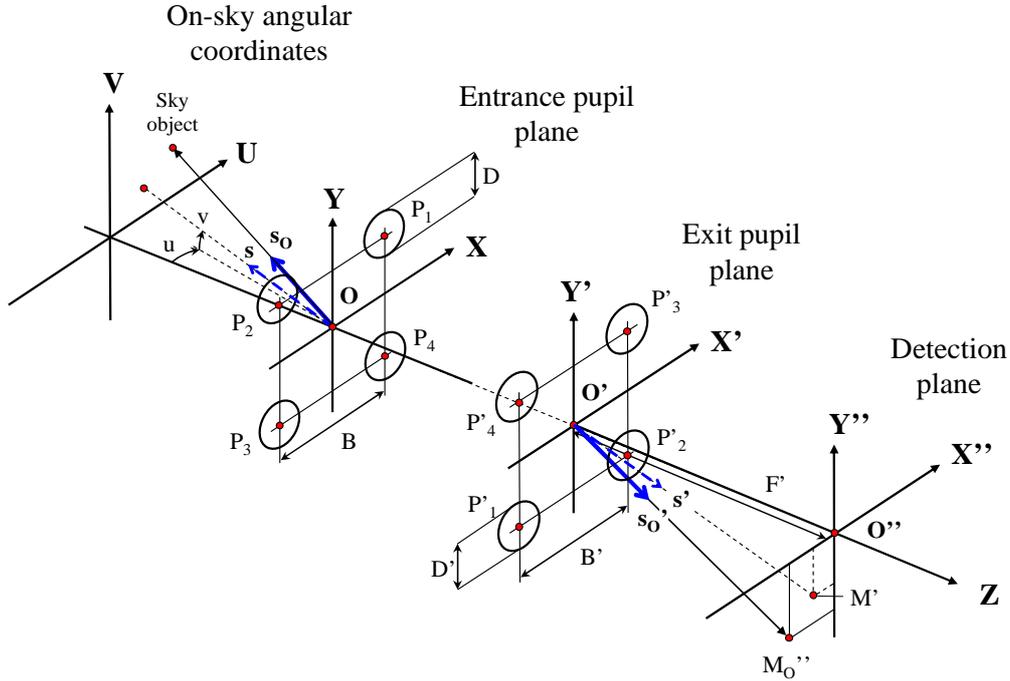

Figure 1: Sketch of the used reference frames on-sky (U,V), on the entrance pupil plane (O,X,Y) and on the exit pupil plane (O',X',Y'). The coordinate system (O",X",Y") attached to the image plane is optically conjugated with the (U,V) reference frame.

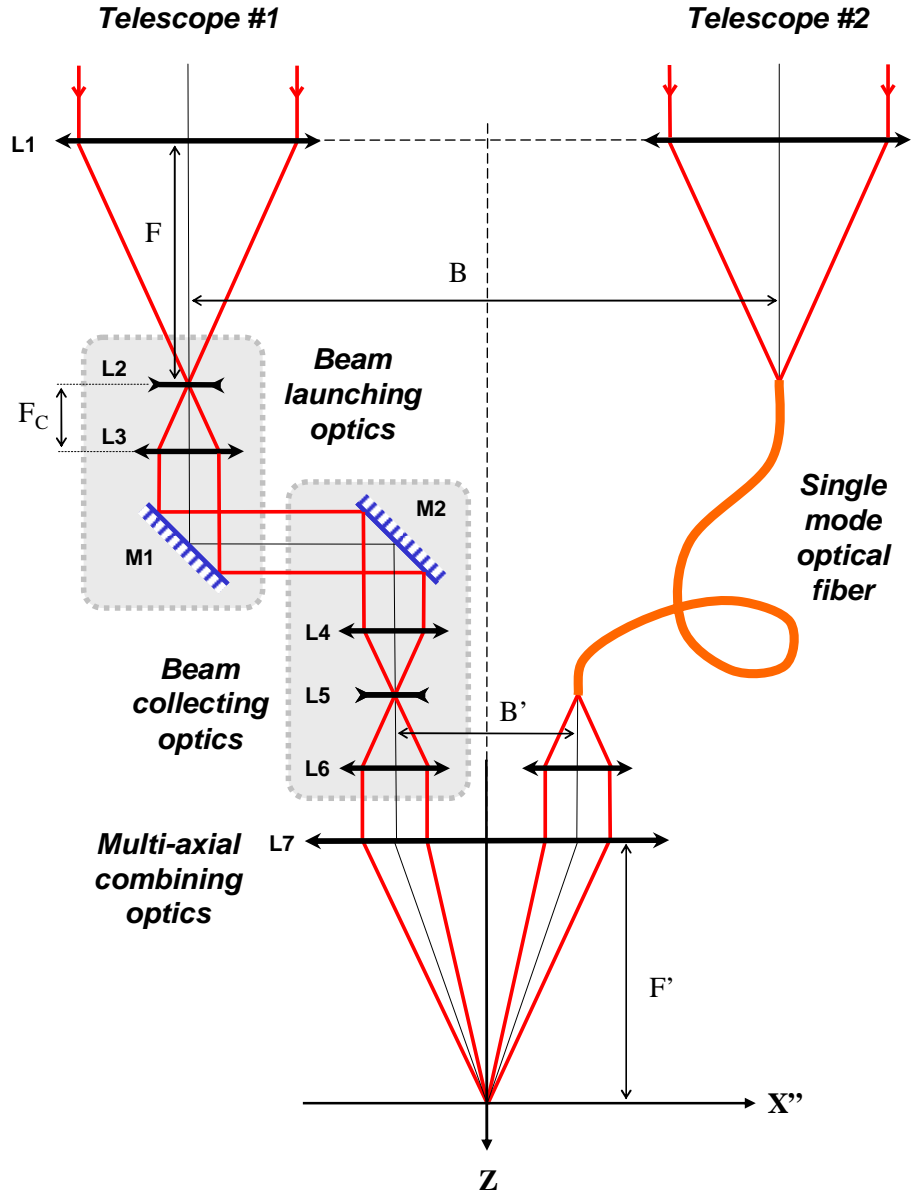

Figure 2: Schematic optical layout of a multi-aperture interferometer, including achromatic phase shifters (APS) used in nulling mode.

### 3     APPLICATION TO FIZEAU INTERFEROMETERS

In this section are discussed the typical Object-Image relationship of multi-aperture Fizeau interferometers (§ 3.1) and the influence of pupil re-imaging aberrations along the interferometer arms on their performance (§ 3.2). The limiting capacity of this type of interferometers for pseudo-imaging planets in nulling mode will be addressed in another section (§ 4.3).

This section is supported by a set of numerical simulations whose numerical parameters are summarized in Table 1 for all considered cases. The numerical values of F and $F_C$ are respectively equal to F = 50 m and $F_C$ = 10 mm, leading to an optical compression factor $m$ = 1/500. All computations were carried out at the wavelength $\lambda$ = 10 µm, which is a typical number for nulling interferometers working in the mid-infrared band.

**Table 1: Numerical values of main physical parameters for all simulated cases.**

| Case | Number of entrance and exit pupils | B (m) | D (m) | B' (mm) | D' (mm) | Section |
|------|------------------------------------|-------|-------|---------|---------|---------|
| Ideal Fizeau interferometer | 8 | 20 | 5 | 40 | 10 | § 3.1 |
| Pupil aberrations | 2 and 8 | 20 | 5 | 40 | 10 | § 3.2 |
| Imaging capacity in nulling mode | 24 | 20 | 1 | 0-8-40 | 2 | § 4.3 |

### 3.1 Theoretical relationships

Classically, Fizeau interferometers present the unique property that their output pupil is the scaled replica of their entrance pupil: all the entrance and exit sub-apertures as well as their relative arrangement are perfectly homothetic. Mathematically this condition implies that:

$$\mathbf{O'P'_n} = m\, \mathbf{OP_n}, \quad (1 \leq n \leq N), \quad (5)$$

also being known as the "Pupil in = Pupil out" condition [4] or "golden rule of interferometry" [5]. In that case, inserting Eq. 5 into Eqs. 1 readily leads to the familiar convolution product of Fourier optics between the object O(**s**) and its image I(**s**) formed by the multi-aperture instrument:

$$I(\mathbf{s}) = [PSF_T(\mathbf{s})\, F(\mathbf{s})] * O(\mathbf{s}), \quad (6a)$$

where function:

$$F(\mathbf{s}) = \left| \sum_{n=1}^{N} a_n \exp[i\varphi_n] \exp[i\, k\, \mathbf{s}\, \mathbf{OP_n}] \right|^2 \quad (6b)$$

was named "Far-field Fringe Function" (FFF) in Ref. [8] because it describes the interference pattern that would be observed in the image plane if all sub-pupils were reduced to a pinhole. It follows from Eq. 6a that Fizeau interferometers possess the natural ability to form real images of an observed sky object, being eventually disturbed by shifted replicas of the same object generated by the FFF. Inserting now the condition 5 into Eqs. 2 and 4 allows retrieving two basic properties of Fizeau interferometers:

1) The generalized PSF of the optical system does not depend any longer on vector $\mathbf{s_O}$, hence the PSF is invariant over the whole interferometer FoV, and equal to $PSF_T(\mathbf{s})\, F(\mathbf{s})$.

2) The maximal achievable FoV of the Fizeau interferometer can be approximated by the very simple relationship:

$$FoV(\mathbf{s}) = \left| \sum_{n=1}^{N} a_n \exp[i\varphi_n] \right|^2. \quad (7)$$

Hence the FoV should be uniform, taking a constant value that only depends on the amplitude transmission factors $a_n$ and phase-shifts $\varphi_n$ of the interferometer. For a classical imaging instrument being perfectly co-phased ($\varphi_n = 0$. whatever is n), FoV(**s**) is uniformly equal to 1, which is in agreement with the golden rule of interferometry. Eq. 7 suggests however that this golden rule may be of dramatic consequence when nulling interferometers are considered, since the phase-shifts $\varphi_n$ generated by their Achromatic Phase Shifters (APS) must be chosen so as to cancel the light originating from the central star. This has the consequence that the invariant PSF is equal to zero at its theoretical center as illustrated in Figure 3, where the case of an eight-telescope Fizeau interferometer is considered in both imaging and nulling modes (the squared arrangement of the array is sketched on the left panel of the Figure, using numerical parameters on the first row of Table 1). Effectively, the computed PSF shows a dark center in nulling mode, and following Eq. 7 one may expect the FoV to be uniformly dark everywhere, therefore making undetectable any extra-solar planet orbiting around its parent star. But here the rigorous expression of the FoV must be evaluated from Eq. 3, turning into the following expression when condition 5 is fulfilled:

$$FoV(\mathbf{s}) = \iint_{\mathbf{s_O} \in \Omega_O} PSF_T(\mathbf{s} - \mathbf{s_O}) \left| \sum_{n=1}^{N} a_n \exp[i\varphi_n] \exp[-i k \mathbf{OP_n}(\mathbf{s} - \mathbf{s_O})] \right|^2 d\Omega_O. \quad (8)$$

From Eq. 8 FoV(**s**) is clearly a constant number that can be evaluated numerically from the parameters of Table 1, and is empirically found to tend toward unity as the B/D ratio becomes larger. Thus in that case, we conclude that a nulling Fizeau interferometer do not fundamentally differ from its imaging version, as will be confirmed by the numerical simulations presented in section 4.3.

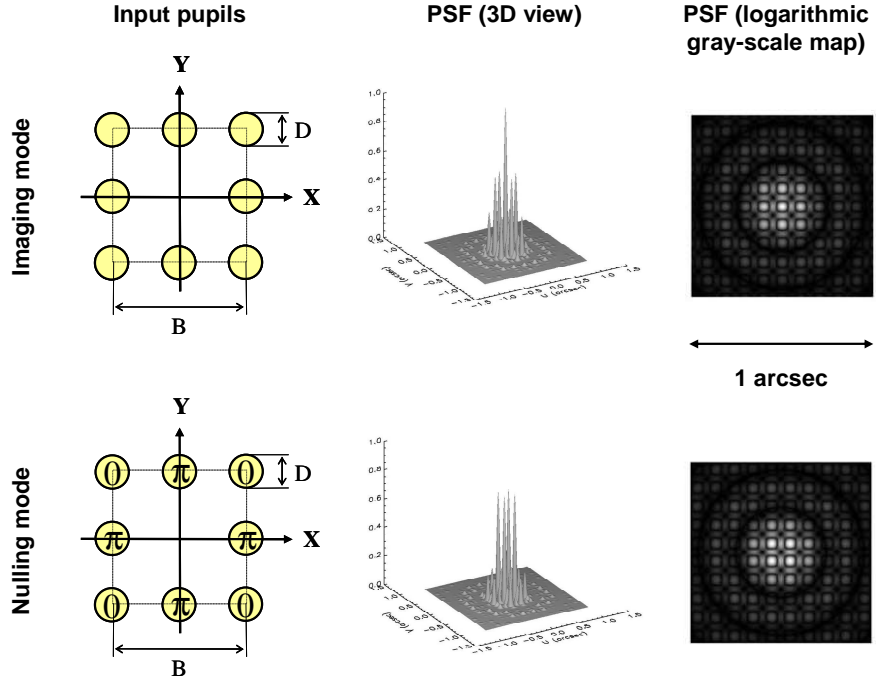

Figure 3: Invariant PSF formed by an eight-telescope Fizeau interferometer in both imaging (top) and nulling modes (bottom). The achromatic phase-shifts $\varphi_n$ of the nulling version are indicated on the left bottom panel.

### 3.2 Deriving pupil imaging requirements

So far was assumed that the angular orientations $(u_O,v_O)$ and $(u,v)$ in vectors $\mathbf{s_O}$ and $\mathbf{s}$ respectively, are small quantities. In the frame of a first-order approximation, this leads to the definition of the general Object-Image relationships of Eqs. 1, and further to its application to the special case of Fizeau interferometers in Eqs. 6 [8]. But indeed, the general relations 1-3 can still be employed whatever are the expressions of the scalar products $\mathbf{s_O}\ \mathbf{OP_n}$ and $\mathbf{s}\ \mathbf{O'P'_n}$. It allows, for example, to study the effects of pupil re-imaging aberrations on the interferometer performance, and first of all of their axial shifts along different arms.

Let us consider a sparse-aperture interferometer whose entrance and exit pupils are affected are suffering from axial shifts $dz_n$' and $dz_n$' ($1 \le n \le N$) with respect to their nominal planes P and P'. All the employed notations are indicated in Figure 4 for the case n = 2 sub-apertures. Entrance pupil shifts $dz_n$ may originate from positioning errors of the collecting telescopes (here assumed to be the stop apertures as on the left side of the Figure), while exit pupil shifts $dz_n$' will be generated by focusing errors in pupil conjugation optics (shown by gray areas in Figure 2). Then both vectors $\mathbf{s_O}$ and $\mathbf{OP_n}$ shall be developed at the second-order:

$$\mathbf{s_O} = \begin{cases} \sin u_0 \approx u_0 \\ \cos u_0 \sin v_0 \approx v_0 \\ \cos u_0 \cos v_0 \approx 1 - u_0^2/2 - v_0^2/2 \end{cases} \quad (9b), \quad \text{and:} \quad \mathbf{OP_n} = \begin{cases} x_n \\ y_n' \\ dz_n \end{cases}. \quad (9b)$$

Similarly, vectors $\mathbf{s}$ and $\mathbf{O'P'_n}$ can be written in this second-order approximation:

$$\mathbf{s} = \begin{cases} \sin u \approx u \\ \cos u \sin v \approx v \\ \cos u \cos v \approx 1 - u^2/2 - v^2/2 \end{cases} \quad (9c), \quad \text{and:} \quad \mathbf{O'P'_n} = \begin{cases} x'_n (1 + dz'_n/F') \\ y'_n (1 + dz'_n/F') \\ dz'_n \end{cases}, \quad (9d)$$

where F' stands for the focal length of combining optics. Then the OPD function $\xi(\mathbf{s_O},\mathbf{s})$ defined in Eq. 1b becomes:

$$\xi(\mathbf{s_O},\mathbf{s}) \approx dz_n - dz'_n/m + u_0 x_n + v_0 y_n - (u\, x'_n + v\, y'_n)(1 + dz'_n/F')/m$$
$$- dz_n (u_0^2 + v_0^2)/2 - dz'_n (u^2 + v^2)/2m. \quad (10)$$

The previous expression of $\xi(\mathbf{s_O},\mathbf{s})$ can be decomposed into three different types of terms, where angles $u_O$, $v_O$, $u$ and $v$ have identical power numbers 0, 1 and 2:

1) The null-order term $dz_n - dz'_n/m$ is a constant OPD, or piston error that is usually compensated for by means of a delay line inserted within the optical train (not shown in Figure 2), hence it can be neglected.

2) A first-order term with respect to $u_O$, $v_O$, $u$ and $v$ is directly related to the geometry of the input and output apertures of the interferometer. In the Fizeau case the "golden rule" of Eq. 5 applies, so that $x'_n = m\, x_n$ and $y'_n = m\, y_n$, therefore the relevant OPD terms cancel. But there remains a residual OPD $(u\, x'_n + v\, y'_n)\, dz'_n/F'$ demonstrating a violation of the golden rule and consequently reducing the maximal FoV achievable by the interferometer. This term will also be neglected, however, since herein we are mainly interested in higher-order effects.

3) Finally, there exists second-order terms proportional to the longitudinal pupil shifts and to the square power of angles $u_O$, $v_O$, $u$ and $v$. This term is responsible for interference patterns distortions, as will be illustrated by the following numerical simulations.

The consequences of pure second-order sub-pupil aberration are illustrated in Figure 5 and Figure 6 for the cases when N = 2 and N = 8 telescopes respectively. Numerical simulations were carried out, based on the physical parameter values given in the second row of Table 1. All the entrance sub-pupil shifts $dz_n$ were set to zero, while the effects of exit pupil shifts being equal to $dz'_n = 0$ mm (ideal case, no pupil aberrations), $dz'_n = 0.05$ mm, and $dz'_n = 0.1$ mm were modeled successively. It must be noted that the two last cases correspond to equivalent displacements of the collecting telescopes of 12.5 and 25 m respectively. These simulations are providing the following outputs:

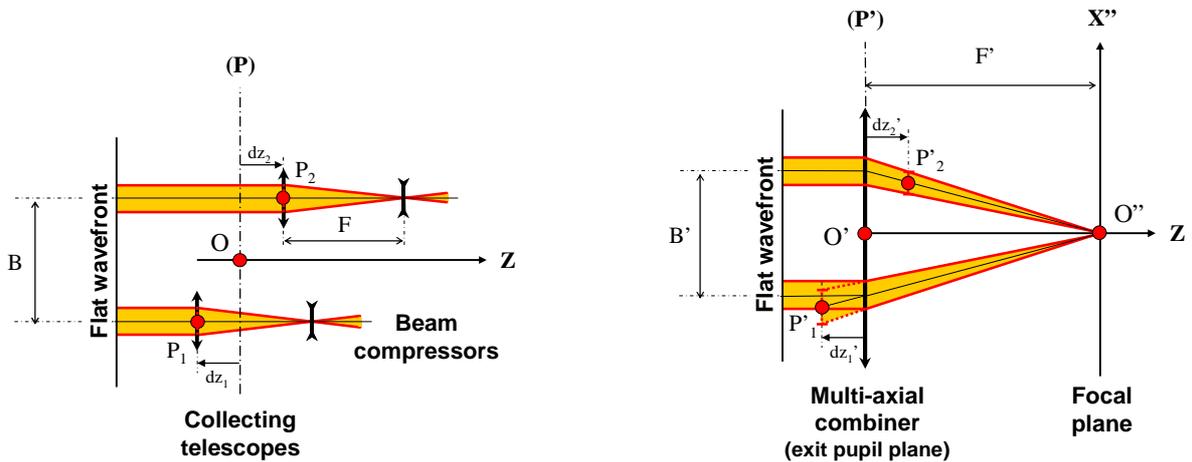

Figure 4: Illustration of sub-aperture axial shifts $dz_1$ and $dz_2$ near the entrance pupil plane P (left side), and $dz_1$' and $dz_2$' near the exit pupil plane P' (right side).

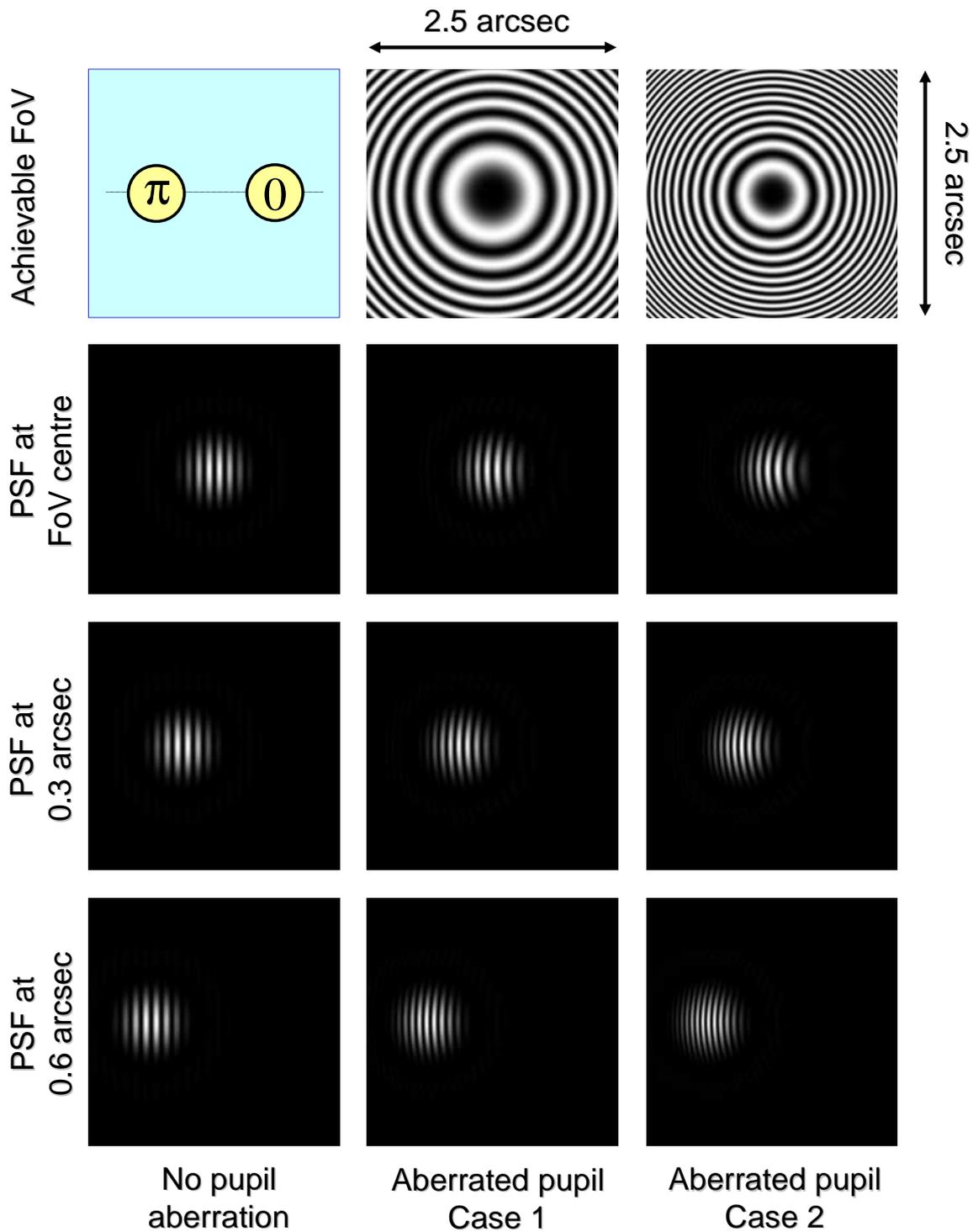

Figure 5: Illustrating the effects of second-order sub-pupils aberration for the case N = 2 telescopes. The sup-pupils arrangement (including phase-shifts) is shown on the upper left corner. In the first row are two gray-scale maps of the achievable FoV for different axial shifts $dz'_n$. From the second to the last rows are displayed the interferometer PSFs at different FoV locations. The left column corresponds to the ideal case when there is no pupil aberration (it can be seen that the PSF shape is invariant), to be compared with the central and right columns where pupil aberration is present.

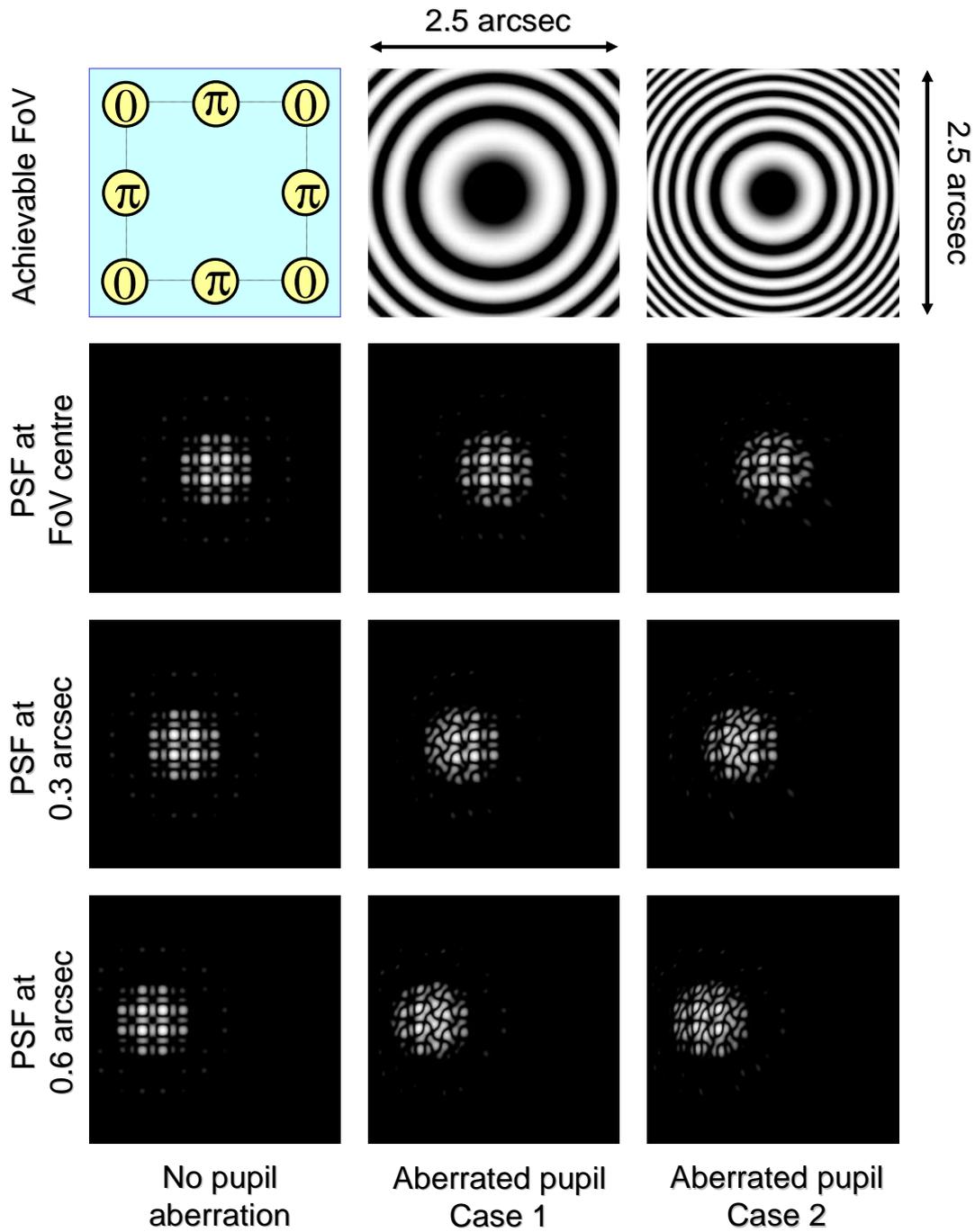

Figure 6: Same graphics presentation than in Figure 5 for the case N = 8 telescopes.

o The PSFs of the interferometer are readily obtained by combining Eqs. 2 and 10. Their gray-scale maps are displayed in the three last rows of Figure 5 and Figure 6 at different FoV locations. It is seen that the fringe patterns are more and more distorted as pupil aberration is increasing (i.e. straight fringes becoming circular). Also of interest is the fact that the PSF is no longer FoV-invariant in presence of pupil aberration, confirming that the golden rule of Fizeau interferometers is not respected due to second-order OPD terms.

- o The maximal achievable FoV is estimated by combining Eqs. 3 and 10. It can be visualized in the first rows of Figure 5 and Figure 6 by the dark central spot, since this is a nulling interferometer. The angular radius of this dark area (turning into a bright one for imaging Fizeau interferometers) is inversely proportional to the amount of pupil aberration, and looks suitable to defining more quantitative optical requirements on the pupil re-imaging quality of the whole system.

One remarkable achievement of the presented formalism and computational method is that no actual Fourier or Fresnel transform algorithms are required explicitly, in opposition with other proposed methods [10]. Hence an appreciable gain in computing time and accuracy shall result. Finally, it is worth mentioning that:

- The here above analytical development for pupil re-imaging imperfections can be applied with no difficulty to other types of interferometers differing for the basic Fizeau layout.

- Further, Eqs. 1-3 are also well suited to introducing variations of the optical compression factor *m* on any type of interferometer[1] (for the Fizeau scheme this implies other deviations from its golden rule). Hence the whole set of Eqs. 1-3 and 10 together provide us with a robust model enabling rapid and efficient evaluation of interferometer performance at system level. This approach is fully complementary with optical design methods involving diverse geometrical aberrations such as described in Ref. [11].

## 4 BEYOND CLASSICAL FIZEAU INTERFEROMETER

Two possible evolutions of nulling Fizeau interferometers hunting for habitable extra-solar planets in the future decades are briefly evoked below, both being based on somehow deliberate violations of the "golden rule" applicable to Fizeau interferometers. They are the Axially Combined Interferometer (§ 4.1), and the crossed-cubes nulling interferometer (§ 4.2). Some of their nulling and pseudo-imaging properties are compared with those of classical Fizeau interferometers in sub-section 4.3.

### 4.1 Axially Combined Interferometer (ACI)

The axially combined interferometer may be considered as a special case of Michelson[2] interferometer where all output sub-pupils are merged together (i.e. $\mathbf{O'P'_n} = \mathbf{0}$ for $1 \leq n \leq N$ in Eqs. 1-4). Given an interferometer constituted of N separated telescopes, this condition can be realized by means of an arrangement of 3N/2 cascaded beamsplitters such as represented in Figure 7. A number of ACIs have already been designed and experimented for imaging purpose [12-13], and they can readily be turned into nulling instruments by means of APS devices as in Bracewell's original concept [6]. The specific Object-Image relationship of the ACI has been demonstrated and discussed in Ref. [8], writing as:

$$I(\mathbf{s}) = PSF_T(\mathbf{s}) * [F(\mathbf{s})\, O(\mathbf{s})], \qquad (11)$$

where the far-field fringe function F(**s**) has the same analytical expression than in Eq. 6b. The latter relationship is quite remarkable, since it implies that the observed sky-object is masked by the FFF of the interferometer array *before* diffraction from the single pupil of the telescope occurs. This is perhaps the fundamental reason why the ACI design is so appealing for nulling interferometry, because it allows in principle to cancel all the light originating from a bright central star, regardless of diffraction effects. Another important consequence of Eq. 11 is that deep nulling should be feasible even with imperfect optics (i.e. $PSF_T(\mathbf{s})$ differing from an ideal Airy distribution), provided that the defects of all telescopes and relay optics are identical along the N interferometer arms. In practice however, this condition should still require the use of spatial or modal wavefront filtering devices located at the image plane of the system, as was foreseen for all the studied projects.

---

[1] Or of its combiner focal length F', should it be made of different segments.
[2] For a discussion about the differences between Fizeau and Michelson interferometers, see e. g. Ref. [9], § 3.

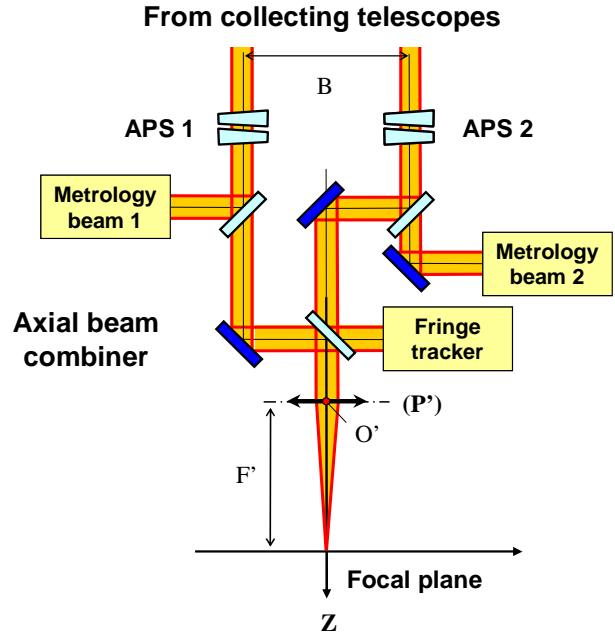

Figure 7: Schematic optical layout of an axially combined interferometer (from Ref. [9]).

### 4.2 Crossed-cubes nulling interferometer

An intermediate configuration between the ACI and the usual Fizeau layout consists in deliberately breaking the golden rule. In that case Eq. 5 is not valid and **O'P'$_n$** = $m'$ **OP$_n$** with $m' \neq m$ as for the historical Michelson's 20" interferometer on Mount Wilson [3] that had two "densified" exit sub-pupils. A good example for such configurations is the Crossed-cubes nulling interferometer (CCN) that is the subject of another communication in this conference [14], to which the reader is invited to refer. One remarkable property of the CCN is that it authorizes pre-determined violations of the homothetic rule, since its entrance and exit baselines B and B' are linked together by the relation (see Figure 8):

$$B' = A\left(1 - \tan\theta\right) - B = A\left(1 - 1/\sqrt{2n^2(\lambda) - 1}\right) - B, \qquad (3)$$

where A is the hypotenuse of the cube, $\theta$ the refracted angle, and $n(\lambda)$ the refractive index of the cube material. In particular, A can be sized so as to achieve maximal densification of the output beams, as illustrated on the right side of the Figure.

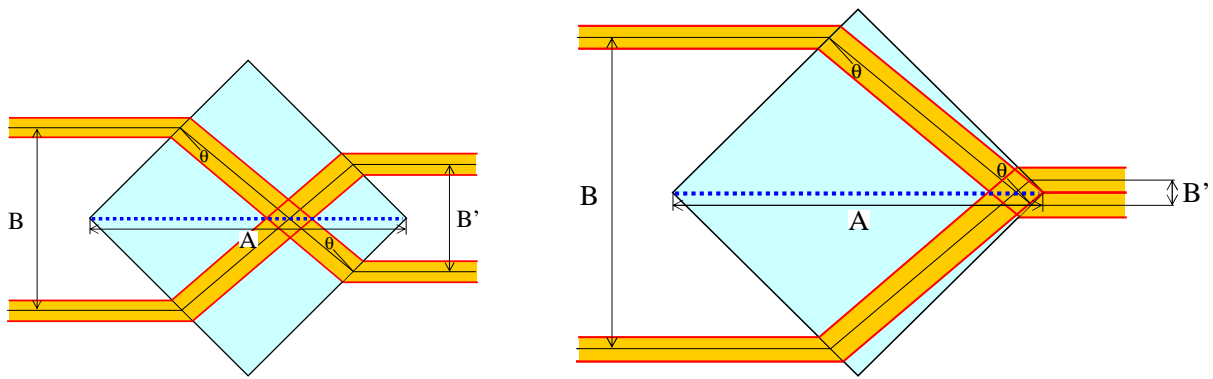

Figure 8: Adjustment of the CCN exit baseline B' as function of the entrance baseline B and cube parameters A and $\theta$.

## 4.3 Imaging properties of nulling interferometer

In this section is finally provided a qualitative interpretation of the imaging properties of nulling interferometers. It has been mentioned in sections 3.1 and 4.1 that the Object-Image relationships applicable to Fizeau and axially combined interferometers are defined by Eqs. 6a and 11 respectively. But it is also of interest to consider the intermediate case of nulling Michelson interferometers (such as the CCN, see § 4.2), for which only Eq. 1a is applicable. The nulling properties of all those instruments employing different combination schemes are illustrated in Figure 9. Here is studied the case of a futuristic 24-telescope nulling interferometer having a checkerboard distribution of its phase-shifts $\varphi_n = 0$ or $\pi$ as indicated on the panel (c). All computations are carried out using the geometrical parameters summarized in Table 1 at the same central wavelength $\lambda = 10$ µm. We consider successively a nulling ACI with a maximal entrance baseline B = 20 m and a null exit baseline B' = 0 mm, then the case when the individual exit sub-pupils have joining edges, corresponding to the maximal achievable densification (B' = 8 mm), and finally a nulling Fizeau interferometer obeying to the golden rule (B' = 40 mm).

For each case is displayed a gray-scale map of the formed image of a fictitious astronomical scene, composed of a bright central star and an off-axis companion of 50 % relative luminosity. Here we obviously aim at nulling the central star and isolating its companion, but the results presented on panels (d), (e) and (f) of Figure 9 evidence clear differences between the three considered combining schemes:

1) Only the nulling ACI demonstrates a full extinction of the central star, as illustrated on Figure 9-d. The produced image appears as a luminous halo fully originating from the companion (and centered on it), enlarged by the diffraction lobe of $PSF_T(\mathbf{s})$ as predicted by Eq. 11.

2) Even for maximal densification, the nulling Michelson interferometer cannot achieve deep extinction of the central star: here the brightest central lobe still originates from the companion, but parasitic images of the star are apparent at the FoV corners. However these replicas remain fainter than the observed companion (here by a factor of 56 %), and it may be assumed that the searched planet can readily be isolated from them.

3) Finally, the image produced by the nulling, homothetic Fizeau interferometer looks quite the same as would be observed with a "constructive" version of it: here the main difference between nulling and imaging modes is that the astronomical scene has been shifted angularly, the central star and its companion having apparently been swapped without any noticeable contrast enhancement. Hence it is concluded that the nulling capacity of this instrument has been lost definitively.

## 5 CONCLUSION

In this paper were reviewed some classical concepts of multi-aperture, imaging and nulling interferometers in the perspective of a first-order Fourier optics formalism. Various topics were revisited and discussed, such as the "golden rule of interferometry", maximal achievable Field of View, performance degradation due to pupil aberration, and the actual imaging capacities of sparse-aperture nulling interferometers. The Object-Image relationships applicable to the presented optical systems have also been illustrated with the help of numerical simulations. The conclusions of this study are that the most suitable combining schemes for nulling purpose seems to be the axially combined interferometer or a CCN-like design with joining exit pupil edges. It has also been confirmed that respecting the classical golden rule of imaging interferometry severely hampers the nulling capacity. Finally, this paper provides the reader with a set of quick computing tools, not requiring any Fourier or Fresnel transform, allowing fast and accurate calculation of the point-spread function, field of view and imaging capacity of these complex high angular resolution systems; in presence of certain types of instrumental defects.

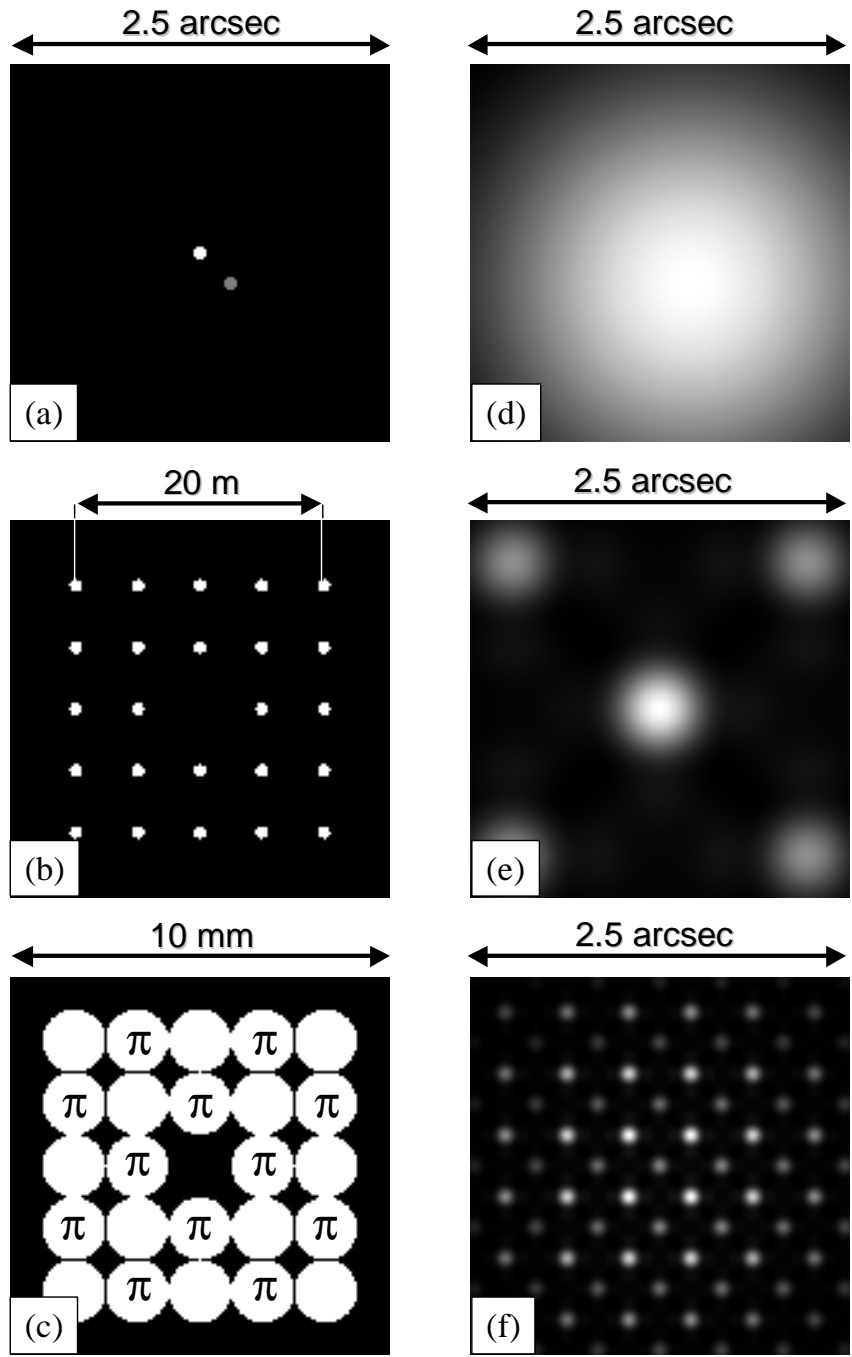

Figure 9: Imaging properties of a 24-telescope nulling interferometer equipped with different combining optics. A fake sky object made of one central star and its half-power companion is shown in panel (a). View (b) depicts the geometry of the interferometer entrance pupil. In (c) are illustrated the checkerboard distribution of the phase-shifts $\varphi_n$ and the geometrical arrangement of the exit sub-pupils for the maximal densification case. Panels (d), (e) and (f) show gray-scale maps of the obtained images, respectively for the cases of axial combination, densified Michelson, and homothetic Fizeau interferometers.

# REFERENCES

[1] H. Fizeau, Rapport sur le concours du Prix Bordin de l'année 1867, Comptes Rendus des Séances de l'Académie des Sciences vol. 66, p. 932-934 (1868).
[2] E. Stéphan, "Sur l'extrême petitesse du diamètre apparent des étoiles fixes," Comptes Rendus des Séances de l'Académie des Sciences vol. 78, p. 1008-1012 (1874).
[3] A. A. Michelson, F. G. Pease, "Measurement of the diameter of alpha Orionis with the interferometer," Astrophys. J. vol. 53, p. 249-259 (1921).
[4] J. M. Beckers, "Field of view considerations for telescope arrays," Proceedings of the SPIE vol. 628, p. 255-260 (1986).
[5] W. A. Traub, "Combining beams from separated telescopes," Applied Optics vol. 25, p. 528-532 (1986).
[6] R. N. Bracewell and R. H. MacPhie, "Searching for non solar planets," Icarus vol. 38, p. 136-147 (1979).
[7] F. Hénault, "Imaging power of multi-fibered nulling telescopes for extra-solar planet characterization," Proceedings of the SPIE vol. 8151, n° 81510A (2011).
[8] F. Hénault, "Simple Fourier optics formalism for high angular resolution systems and nulling interferometry," JOSA A vol. 27, p. 435-449 (2010).
[9] F. Hénault, "PSF and field of view characteristics of imaging and nulling interferometers," Proceedings of the SPIE vol. 7734, n° 773419 (2010).
[10] D Mekarnia, J Gay, "Altération du facteur de visibilité par diffraction de Fresnel en synthèse d'ouverture," Journal of Optics vol. 20, p. 131-140 (1989).
[11] E. E. Sabatke, J. H. Burge, P. Hinz, "Optical design of interferometric telescopes with wide fields of view," Applied Optics vol. 45, p. 8026-8035 (2006).
[12] J. E. Baldwin, R. C. Boysen, G. Cox, C. A. Haniff, J. Rogers, P. J. Warner, D. M. A. Wilson, C. D. Mackay, "Design and performance of COAST," Proceedings of the SPIE vol. 2200, p. 118-128 (1994).
[13] M. M. Colavita, J. K. Wallace, B. E. Hines, Y. Gursel, F. Malbet, D. L. Palmer, X. P. Pan, M. Shao, J. W. Yu, A. F. Boden, P. J. Dumont, J. Gubler, C. D. Koresko, S. R. Kulkarni, B. F. Lane, D. W. Mobley, G. T. van Belle, "The Palomar Testbed Interferometer," Astrophysical Journal vol. 510, p. 505-521 (1999).
[14] F. Hénault, A. Spang, "Cheapest nuller in the world: crossed beamsplitter cubes," Proceedings of the SPIE vol. 9146 [This conference, Ref. 9146-90].